\documentclass[twocolumn,aps,prl,10pt,superscriptaddress,longbibliography,floatfix,citeautoscript]{revtex4-2}
\usepackage{amsmath}
\usepackage{amssymb}
\usepackage{bm}
\usepackage{glossaries}
\usepackage{graphicx}
\usepackage{multirow}
\usepackage{fancyvrb}
\usepackage[linesnumbered,ruled]{algorithm2e}
\usepackage{siunitx}
\usepackage{subfigure}
\usepackage{threeparttable}
\usepackage{adjustbox}
\usepackage{textcomp}

\usepackage[usenames,dvipsnames,svgnames]{xcolor}
\usepackage{hyperref}  % must come after xr-hyper
\hypersetup{
pdfnewwindow=true,      % links in new window
colorlinks=true,        % false: boxed links; true: colored links
linkcolor=Blue,         % color of internal links
citecolor=Blue,         % color of links to bibliography
filecolor=Blue,         % color of file links
urlcolor=Blue           % color of external links
}

\newacronym{dft}{DFT}{density-functional theory}
\newacronym{eam}{EAM}{embedded-atom method}
\newacronym{dp}{DP}{deep potential}
\newacronym{md}{MD}{molecular dynamics}
\newacronym{bcc}{BCC}{body-centered-cubic}
\newacronym{fcc}{FCC}{face-centered-cubic}
\newacronym{hcp}{HCP}{hexagonal close-packed}
\newacronym{bop}{BOP}{bond-order potentials}
\newacronym{nep}{NEP}{neuroevolution potential}
\newacronym{tis}{TIS}{tetrahedral interstitial site}
\newacronym{ois}{OIS}{octahedral interstitial site}
\newacronym{gsf}{GSF}{generalized stacking fault}
\newacronym{sia}{SIA}{self-interstitial atom}
\newacronym{zpe}{ZPE}{zero-point-energy}
\newacronym{rdf}{RDF}{radial distribution function}
\newacronym{cna}{CNA}{common neighbor analysis}
\newacronym{mlp}{MLP}{machine-learned potentials}
\newacronym{aimd}{AIMD}{\emph{ab initio} molecular dynamics}
\newacronym{rmse}{RMSE}{root-mean-square error}
\newacronym{zbl}{ZBL}{Ziegler-Biersack-Littmark}
\newacronym{gb}{GB}{grain boundary}
\newacronym{nn}{NN}{neural network}

\DeclareSIUnit\angstrom{\text{Å}}
\DeclareSIUnit\atom{\text{atom}}

\begin{document}

\title{Atomistic understanding of hydrogen bubble-induced embrittlement in tungsten enabled by machine learning molecular dynamics}

\author{Yu Bao}
 \thanks{These authors contributed equally to this work.}
\affiliation{Department of Physics, University of Science and Technology Beijing, Beijing, 100083, China}

\author{Keke Song}
 \thanks{These authors contributed equally to this work.}
\affiliation{College of Physics and Information Engineering, Fuzhou University, Fuzhou 350108, China}

\author{Jiahui Liu}
\affiliation{Beijing Advanced Innovation Center for Materials Genome Engineering, University of Science and Technology Beijing, Beijing 100083, China}

\author{Yanzhou Wang}
\affiliation{Department of Applied Physics, P.O. Box 15600, Aalto University, FI-00076 Aalto, Espoo, Finland}

\author{Yifei Ning}
\affiliation{Department of Physics, University of Science and Technology Beijing, Beijing, 100083, China}

\author{Penghua~Ying}
\email{hityingph@tauex.tau.ac.il}
\affiliation{Department of Physical Chemistry, School of Chemistry, Tel Aviv University, Tel Aviv, 6997801, Israel}

\author{Ping Qian}
\email{qianping@ustb.edu.cn}
\affiliation{Department of Physics, University of Science and Technology Beijing, Beijing, 100083, China}

\date{\today}

\begin{abstract}
Hydrogen bubble formation within nanoscale voids is a critical mechanism underlying the embrittlement of metallic materials, yet its atomistic origins remains elusive. Here, we present an accurate and transferable machine-learned potential (MLP) for the tungsten-hydrogen binary system within the neuroevolution potential (NEP) framework, trained through active learning on extensive density functional theory data. The developed NEP-WH model reproduces a wide range of lattice and defect properties in tungsten systems, as well as hydrogen solubility, with near first-principles accuracy, while retaining the efficiency of empirical potentials. Crucially, it is the first MLP capable of capturing hydrogen trapping and H\textsubscript{2} formation in nanovoids, with quantitative fidelity. Large-scale machine-learning molecular dynamics simulations reveal a distinct aggregation pathway where planar hydrogen clusters nucleate and grow along \{100\} planes near voids, with hexagonal close-packed structures emerging at their intersections. Under uniaxial tension, these aggregates promote bubble fracture and the development of regular \{100\} cracks, suppressing dislocation activity and resulting in brittle fracture behavior. This work provides detailed atomistic insights into hydrogen bubble evolution and fracture in nanovoids, enables predictive modeling of structural degradation in extreme environments, and advances fundamental understanding of hydrogen-induced damage in structural metals. 
\end{abstract}

\maketitle
\vspace{0.5cm}

\noindent\textbf{INTRODUCTION}
\noindent 
Hydrogen-induced degradation has long been a critical challenge in materials science, often leading to premature or catastrophic failure of structural metals~\cite{johnson1875some}. Over the past decades, multiple mechanisms have been proposed to explain hydrogen embrittlement~\cite{robertson2015hydrogen}, with the formation and evolution of internal high-pressure bubbles or voids recognized as a central factor~\cite{tetelman1963direct,neeraj2012hydrogen,xie2015situ,hou2019predictive}. 
Tungsten (W) is a representative system for studying hydrogen-induced embrittlement, as it is widely considered a leading candidate for plasma-facing components in future fusion reactors owing to its exceptional physical properties~\cite{pitts2013full, hirai2014iter}. As a typical metal exposed to hydrogen-rich environments, tungsten has exhibited extensive hydrogen bubble and failure phenomena in experiments~\cite{balden2014deuterium, zhao2018investigation, guo2018edge, chen2019irradiation, chen2020nucleation, chen2020growth}. Despite extensive experimental evidence of hydrogen bubble–induced embrittlement~\cite{guo2018edge, chen2020nucleation, chen2020growth}, the atomic-scale processes governing bubble nucleation and growth remain poorly understood, due to the challenges of in situ characterization of their internal structure and pressure.

Atomistic simulations provide a powerful complement to experiments. First-principles calculations at the \gls{dft} level can accurately describe hydrogen interactions with vacancies, \glspl{gb}, surfaces, and dislocations at zero temperature~\cite{liu2009vacancy,ismer2009interactions,tatayama2003stability,geng2001influence,zhou2016chemomechanical,davide2015first, xiong2023hydrogen,lisa2013ab}. However, their prohibitive cost in both system size and timescale prevents direct resolution of the mechanisms responsible for hydrogen embrittlement. In contrast, \gls{md} simulations based on empirical interatomic potentials can access much larger systems and extended timescales, and have yielded valuable insights into hydrogen embrittlement\cite{song2013atomic,tehranchi2020decohesion,hou2020hydrogen}. However, their reliability is fundamentally limited by the quality of the underlying potentials, which show large discrepancies in predicting H trapping at defect\cite{wang2017embedded, mary2015comparison, starikov2022angular} and fail to capture H-nanovoids interactions or reproduce experimentally observed gigapascal-level internal H\textsubscript{2} \cite{hayward2013interplay,wang2017embedded}. These deficiencies cast serious doubt on their ability to predict key processes such as hydrogen bubble nucleation and evolution.

By learning directly from quantum-mechanical electronic structure data, \glspl{mlp} \cite{behler2007generalized, bartok2010gaussian, shapeev2016moment, wang2018deepmd, fan2021neuroevolution} have emerged over the past two decades as a powerful framework that overcomes long-standing accuracy-efficiency trade-off of \gls{dft} and empirical potentials, enabling large-scale, high-fidelity simulations of complex materials at tractable computational cost \cite{deringer2019machine, friederich2021machine, ying2025advances}. For the W-H system, Wang \textit{et al.} developed a \gls{dp}-based model (denoted as \gls{dp}-WH hereafter)~\cite{wang2023deep} that successfully reproduces the pure W properties and solute H behavior. As the training dataset did not include configurations involving multiple vacancies decorated with numerous H atoms, the model's transferability to H bubble dynamics remains uncertain~\cite{wang2023deep}. 

In this study, we develop a unified \gls{mlp} model for the W-H binary system based on the \gls{nep} framework (hereafter denoted as \gls{nep}-WH)~\cite{fan2021neuroevolution}, using a feedforward \gls{nn} trained against \gls{dft} data via an active learning scheme. The \gls{nep} framework is selected for its combination of near-\gls{dft} accuracy and exceptional computational efficiency~\cite{song2024general}, enabling simulations of more than two million atoms over time spans exceeding \SI{10}{\nano\second}. Such capability is essential for resolving atomistic processes of hydrogen bubble-induced embrittlement. The developed \gls{nep}-WH model surpasses all previously reported  W-H potentials, accurately reproducing fundamental lattice and defect properties of W, solute H behavior in both perfect and defect matrices, and the formation energy of the H$_\textrm{2}$. Crucially, it provides substantial improvements in describing void-hydrogen interactions. Leveraging this model, we perform extensive \gls{md} simulations across nanovoids of varying radii, which reveal the formation of \{100\} planar hydrogen clusters and H-rich \gls{hcp} regions driven by planar hydrogen agglomeration. This evolution gives rise to H concentration-dependent a ductile-to-brittle transition under uniaxial tension.

\begin{figure*}[!] 
\centering 
\includegraphics[width=2\columnwidth]{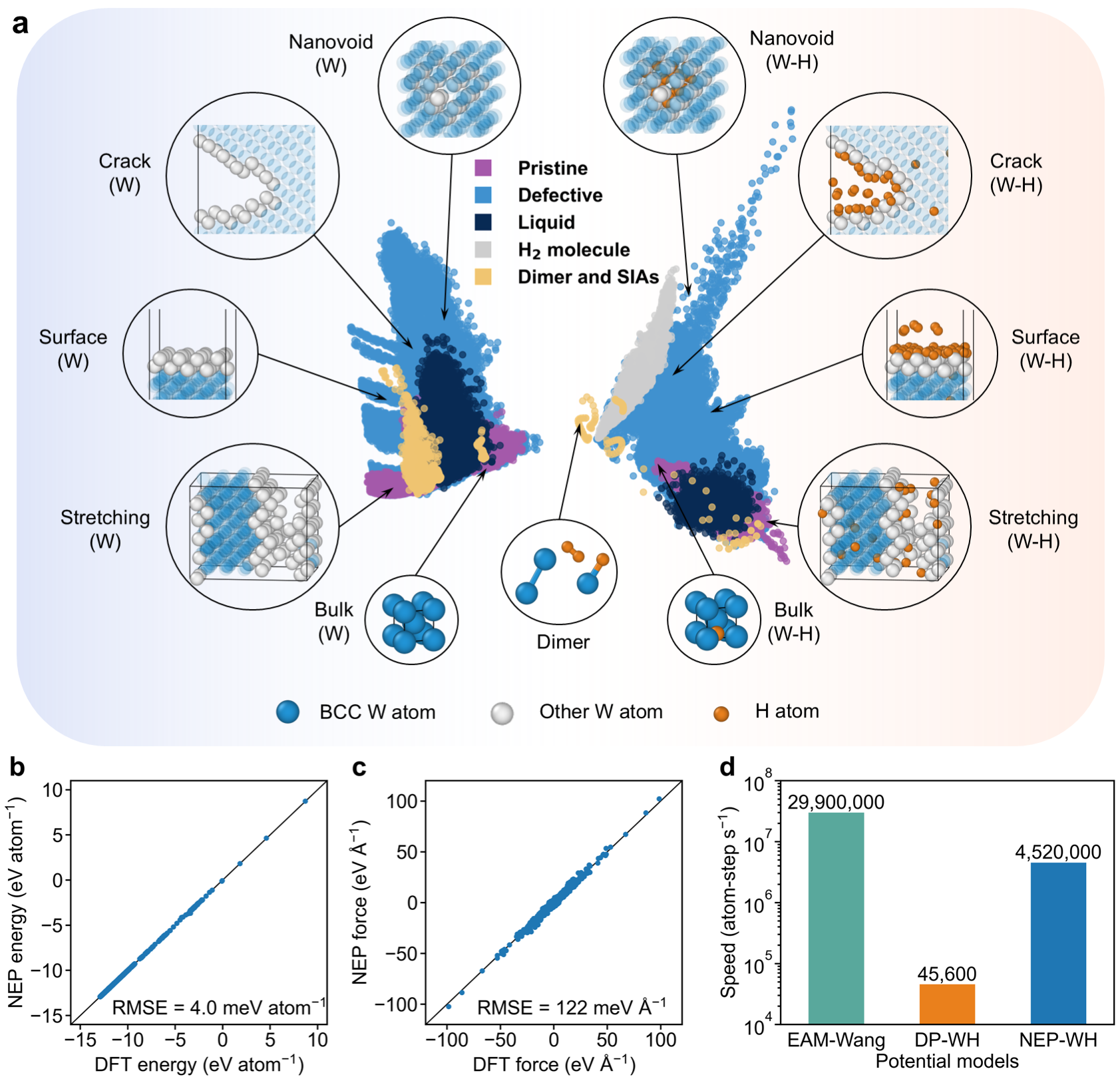}
\caption{
\textbf{Dataset and performance of the \gls{nep}-WH model.} 
\textbf{a} Sketch-map visualization of the training dataset comprising 57,400 structures. Each colored dot represents an individual atomic configuration projected onto a 2D principal component space of the learned \gls{nep} descriptors, constructed using the \textsc{calorine} package \cite{lindgren2024calorine}. The left half shows pure W configurations,including pristine, defective, liquid, dimer, and \gls{sia} structures, while the right half displays their counterparts in the W-H binary system and H\textsubscript{2} molecule. Representative configurations, such as bulk, stretching, surface, crack, and nanovoid structures, sampled from different regions of the map highlight the diversity of the dataset. W atoms identified as \gls{bcc} and other by the \gls{cna} method \cite{Faken1994Systematic} are shown as large blue and white spheres, respectively, and hydrogen atoms are shown as orange spheres. 
\textbf{b} Energy and \textbf{c} force predictions from the \gls{nep}-WH model versus \gls{dft} reference values for the training set. The solid line denotes the identity line, used to guide the eyes. 
\textbf{d} Computational speeds for pure W systems using the \gls{eam}-Wang \cite{wang2017embedded}, \gls{dp}-WH \cite{wang2023deep}, and \gls{nep}-WH model, evaluated on a single NVIDIA V100 GPU with 32 GB of memory.} 
\label{fig:trainingRMSE}
\end{figure*}

\vspace{0.5cm}

\noindent\textbf{RESULTS}

\noindent\textbf{NEP-WH model training}

\noindent We employed an active learning strategy based on committee uncertainty estimates~\cite{Christoph2021Machine, song2024general} from an ensemble of \gls{nep} models to systematically explore the configuration space for training the \gls{nep}-WH model. The initial dataset comprised diverse configurations, including bulk W with and without interstitial H, H-decorated vacancy clusters, manually perturbed structures, and structures sampled from \gls{aimd} simulations. Single-point \gls{dft} calculations using the Perdew-Wang (PW91) functional~\cite{perdew1992atoms} (see Methods for details) were performed to obtain total energies and atomic forces for these structures, which were then used to train an ensemble of four \gls{nep} models. One model from this ensemble was then used to carry out \gls{md} simulations to further sample the configuration space. Model uncertainty was estimated from the variance in predicted forces across the ensemble, and configurations with large discrepancies were identified as outliers. These outliers were recalculated with \gls{dft} calculations to obtain energies, forces, and virial tensors, and subsequently added to the training set for the next iteration. The iterative process continued until the prediction errors across all sampled configurations converged within acceptable limits. The active learning procedure spanned a wide range of thermodynamic conditions, with temperatures from \SI{50}{\kelvin} to \SI{5100}{\kelvin} and pressures from \SI{-2}{\mega\pascal} to \SI{20}{\giga\pascal}. To further enhance robustness, we manually incorporated configurations of melted and strained structures. The final training set included pristine, defective, liquid, and short-range interaction configurations for both W and W-H systems as well as  $\mathrm{H}_2$ molecule, totaling 57,400 structures (8,267,784 atoms) (see \autoref{fig:trainingRMSE}a). Details of the iterative process for different structural types are provided in Supplementary Note S1 and Table S1.

In each iteration, the fourth-generation of \gls{nep} framework \cite{fan2021neuroevolution,song2024general} was employed, with a radial cutoff of \SI{6}{\angstrom} and an angular cutoff of \SI{5}{\angstrom}. To prevent unphysical clustering due to insufficient short-range repulsion, the \gls{zbl} potential \cite{ziegler1985stopping, liu2023large} was integrated into the trained \gls{nep}-WH model. The input file used for training the \gls{nep}-WH model is provided in Supplementary Note S3, and additional details are described in the Methods section.

\begin{figure*}
    \centering
    \includegraphics[width=2\columnwidth]{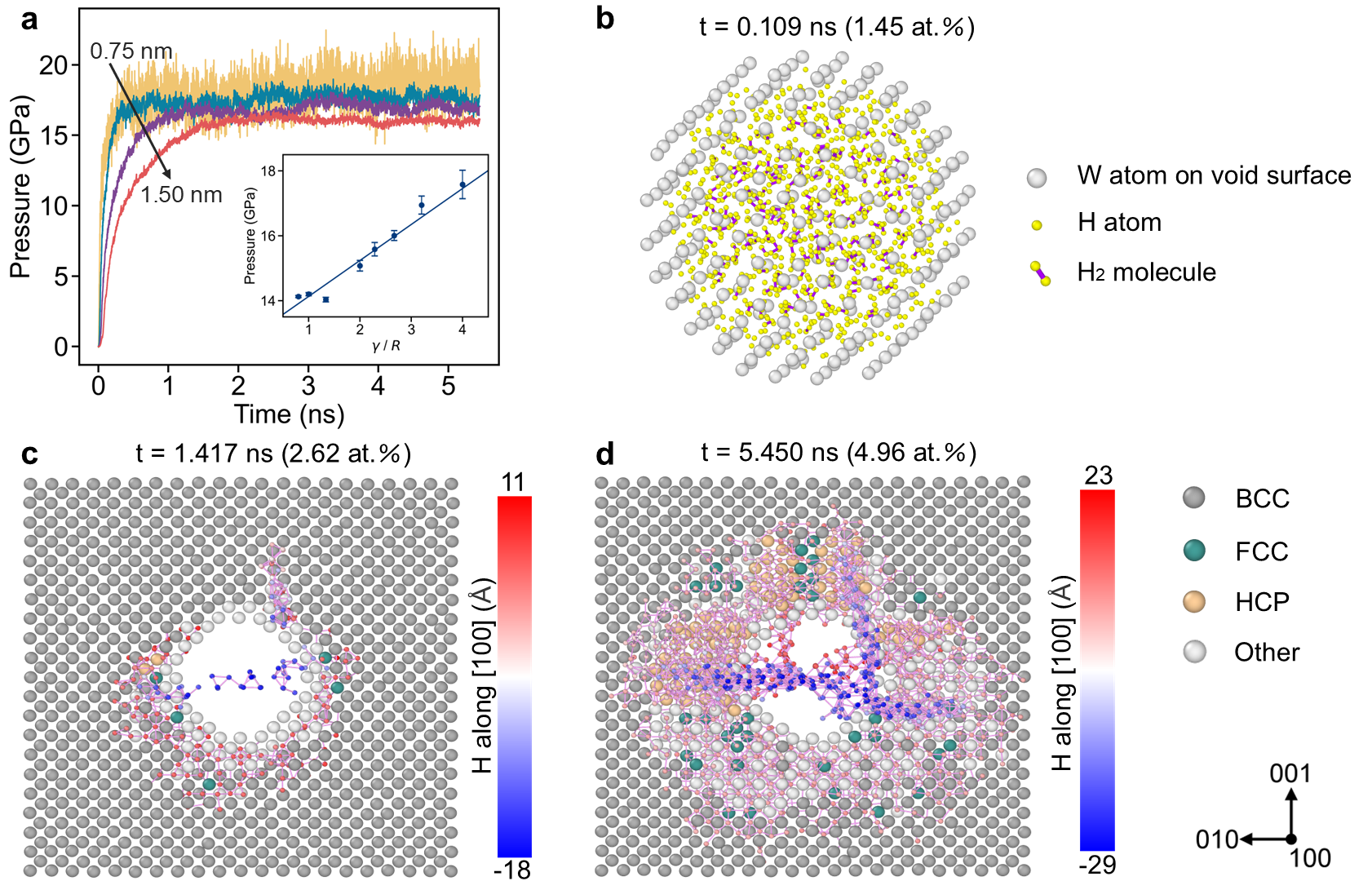}
    \caption{\textbf{Hydrogen solubility behavior in nanovoids}. \textbf{a} Bubble pressure as a function of simulation time for nanovoids with radii of \SI{0.75}{\nano\meter}, \SI{1}{\nano\meter}, \SI{1.25}{\nano\meter}, and \SI{1.5}{\nano\meter} at \SI{600}{\kelvin}. The inset in panel (\textbf{a}) illustrates the relationship between pressure and the surface energy divided by the radius. \textbf{b-d} Snapshots of a \SI{1}{\nano\meter} radius nanovoid at (\textbf{b}) \SI{0.109}{\nano\second}, (\textbf{c}) \SI{1.417}{\nano\second}, and (\textbf{d}) \SI{5.450}{\nano\second}, corresponding to 474, 855, and 1619 H$_\textrm{2}$ molecule insertion, respectively. 
    Panel (\textbf{b}) present a magnified three-dimensional view of the \SI{1}{\nano\meter} nanovoid at \SI{0.109}{\nano\second} with \gls{bcc} atoms removed. Yellow spheres represent hydrogen atoms, while purple bonds indicate H$_\textrm{2}$ molecule. 
    Panels (\textbf{c}) and (\textbf{d}) present [100] views of locally enlarged regions of the \SI{1}{\nano\meter} nanovoid at \SI{1.417}{\nano\second} and \SI{5.450}{\nano\second}, respectively. The local view includes a \SI{65}{\angstrom} thick W slab in the [010] and [001] directions (\SI{\pm32.5}{\angstrom} from the void center), and a \SI{4}{\angstrom} thick slab in [100] near the (100) planer hydrogen cluster. 
    H atoms within a 2.8 \AA\ cutoff radius are considered part of a cluster (indicated by pink bonds). The H atoms in planar clusters are color-coded based on their position along the [100] axis, ranging from blue (minimum coordinate) to red (maximum coordinate). Note that H atoms located inside the void (defined as a sphere with a radius equal to the nanovoid radius plus W-W bond length of \SI{0.28}{\nano\meter}) or identified as isolated H atoms (having fewer than 10 atoms in a cluster) are excluded. 
    W atoms identified by the \gls{cna} method as \gls{bcc}, \gls{fcc}, \gls{hcp}, and Other are shown as large gray, green, orange, and white spheres, respectively.} 
    \label{fig:bubble}
\end{figure*}

The parity plots for energy and force demonstrate the high
accuracy of the \gls{nep}-WH model (\autoref{fig:trainingRMSE}b-c). Despite the wide ranges in both quantities, the achieved \glspl{rmse} are \SI{4}{\milli\electronvolt\per\atom} for energy and \SI{122}{\milli\electronvolt\per\angstrom} for force. In addition to its accuracy, the \gls{nep}-WH model, as implemented in \textsc{GPUMD}~\cite{xu2025gpumd}, exhibits excellent computational performance. It achieves a computational speed of $4.5 \times 10^{6}$ atom step s$^{-1}$, approximately two orders of magnitude faster than the \gls{dp}-WH model~\cite{wang2023deep} implemented in LAMMPS (version 2Aug2023)~\cite{thompson2022lammps}, and only several-fold slower than the \gls{eam} potential by Wang \textit{et al.} (denoted as \gls{eam}-Wang hereafter)~\cite{wang2017embedded}, also implemented in \textsc{LAMMPS} (version 29Aug2024)~\cite{thompson2022lammps} (\autoref{fig:trainingRMSE}(d)). This performance benchmark was obtained by running \gls{md} simulations in the micro-canonical ensemble for 1,000 steps on bulk \gls{bcc} W containing 18,522 atoms, using a single NVIDIA V100 GPU with 32 GB of memory. Notably, this is the maximum system size that the \gls{dp}-WH model can handle under these conditions without encountering memory errors. In contrast, the \gls{nep}-WH model can scale up to five million atoms on the same hardware (see Supplementary Fig.~S1). The superior computational efficiency of \gls{nep}-WH is crucial for subsequent hydrogen bubble simulations, which require extensive sampling to insert H$_\textrm{2}$ molecules until steady-state bubble pressure is reached (see \autoref{fig:bubble}(a)).

\begin{table*}[htbp]
\caption{Static properties of element W and H, and binary W-H systems predicted by \glspl{bop} \cite{juslin2005analytical, li2011modified},  \gls{eam} potentials \cite{wang2017embedded, mason2023empirical}, \gls{dp}-WH \cite{wang2023deep}, and \gls{nep}-WH models as compared to experimental (Exp.) and/or \gls{dft} reference values. For \gls{dft} results, both values from previous studies (Previous \gls{dft}) and those obtained in the present work (\gls{dft}) are included. All lattice parameters $a$ and bond length $L_{\rm b}$ are given in $\rm \AA$, elastic constants $C_{ij}$ in GPa, and all other energies $E$ in eV, expect for surface formation energy $E_{\rm f}^{\rm surf}$, which is given in J m$^{-2}$.
$a_{\rm bcc}$ and $a_{\rm fcc}$ denote the lattice parameters of \gls{bcc} and \gls{fcc} W, respectively. 
$C_{ij}$ denote the elastic constants of \gls{bcc} W. 
$E_{\rm f}^{\rm SIA}$ refers to the formation energy of interstitial dumbbells with different orientations. 
$E_{\rm f}^{\rm vac}$ and $E_{\rm m}^{\rm vac}$ represent the formation energy and migration energy of a mono-vacancy, respectively. 
$E_{\rm b}^{\rm divac-1NN}$ and $E_{\rm b}^{\rm divac-2NN}$ are the binding energies for the first- and second-nearest neighbor divacancy configurations, respectively. 
$E_{\rm f}^{\rm surf}$ denote formation energies of low-index free surfaces with different orientations. 
$E_{\rm f}^{\rm tis}$ and $E_{\rm f}^{\rm ois}$ represent the solution energies of a hydrogen atom at \gls{tis} and \gls{ois}, respectively. 
$E_{\rm m}^{\rm t \rightarrow t}$ denotes H migration energy between two nearest-neighbor \glspl{tis}. 
$E_{\rm b}^{\rm Tplanar}$, $E_{\rm b}^{\rm Oplanar}$, and $E_{\rm b}^{\rm rock-salt}$ denote the binding energies of H in the \gls{tis} and \gls{ois} planar self-clusters, and the rock-salt structures, respectively. 
$E_{\rm b}^{\rm surf}$ are the binding energies of a hydrogen atom on the most stable adsorption site of low-index surfaces. 
$E_{\rm f}^{\rm H_2}$ denotes the formation energy of a H\textsubscript{2} molecule in vacuum. 
$L_{\rm b}^{\rm H_{\rm 2}}$ denotes the bond length of H\textsubscript{2} molecule.
Values taken from previous studies are cited with references; all other values are obtained in the present work.
} % title name of the table  
\centering % centering table

\scalebox{1}{
\begin{tabular}{llllllllll} 
\hline\hline   
 Potential & \gls{bop}-Juslin & \gls{bop}-Li & \gls{eam}-Wang & \gls{eam}-Mason & \gls{dp}-WH & Exp. & Previous \gls{dft} & \gls{dft} & \gls{nep}-WH
\\[0.8ex]  
\hline   
% Entering nst row  
Element W: & & & & & & & & & \\
$a_{\rm bcc}$ & 3.165 \cite{juslin2005analytical} & 3.165 \cite{li2011modified} & 3.140 \cite{wang2017embedded} & 3.145 \cite{mason2023empirical} & 3.186 \cite{wang2023deep} & 3.165 \cite{lide2004crc} & 3.165 \cite{marinica2013interatomic}, 3.172 \cite{szlachta2014accuracy} & 3.175 & 3.176 \\
$a_{\rm fcc}$ & 4.005 \cite{juslin2005analytical} & 4.002 \cite{li2011modified} & 4.053 \cite{wang2017embedded} & 3.975 & 4.040 &  & 4.054 \cite{marinica2013interatomic}, 4.023 \cite{wang2022tungsten} & 4.028 & 4.029 \\
$C_{11}$ & 542 \cite{li2011modified} & 515 \cite{li2011modified} & 544 \cite{wang2017embedded} & 513 \cite{mason2023empirical} & 516 \cite{wang2023deep} & 522 \cite{lide2004crc} & 523 \cite{marinica2013interatomic}, 517 \cite{szlachta2014accuracy} & - & 534 \\ 
$C_{12}$ & 191 \cite{li2011modified} & 203 \cite{li2011modified} & 208 \cite{wang2017embedded} & 201 \cite{mason2023empirical} & 201 \cite{wang2023deep} & 204 \cite{lide2004crc} & 203 \cite{marinica2013interatomic}, 198 \cite{szlachta2014accuracy} & - & 199 \\ 
$C_{44}$ & 162 \cite{li2011modified} & 162 \cite{li2011modified} & 160 \cite{wang2017embedded} & 163 \cite{mason2023empirical} & 146 \cite{wang2023deep} & 161 \cite{lide2004crc} & 160 \cite{marinica2013interatomic}, 142 \cite{szlachta2014accuracy} & - & 156 \\ 
$E_{\rm f}^{\rm SIA-\left \langle111\right \rangle}$ & 9.62 \cite{li2011modified} & 9.33 \cite{li2011modified} & 10.52 \cite{wang2017embedded} & 9.74 \cite{mason2023empirical} & 10.16 \cite{wang2023deep} &  & 10.53 \cite{marinica2013interatomic}, 9.55 \cite{nguyen2006self} & 9.99 & 10.21 \\ 
$E_{\rm f}^{\rm SIA-\left \langle110\right \rangle}$ & 8.77 \cite{li2011modified} & 9.53 \cite{li2011modified} & 10.82 \cite{wang2017embedded} & 9.62 & 10.35 \cite{wang2023deep} &  & 10.82 \cite{marinica2013interatomic}, 9.84 \cite{nguyen2006self} & 10.27 & 10.49 \\
$E_{\rm f}^{\rm SIA-tetra}$ & 8.60 \cite{li2011modified} & 10.75 \cite{li2011modified} & 11.94 \cite{wang2017embedded} & 10.60 & 11.81 &  & 12.27\cite{marinica2013interatomic}, 11.05 \cite{nguyen2006self} & 11.47 & 10.95 \\
$E_{\rm f}^{\rm SIA-\left \langle100\right \rangle}$ & 8.93 \cite{li2011modified} & 12.01 \cite{li2011modified} & 12.86 \cite{wang2017embedded} & 9.97 \cite{mason2023empirical} & 12.03 \cite{wang2023deep} &  & 12.87 \cite{marinica2013interatomic}, 11.49 \cite{nguyen2006self} & 11.96 & 11.07 \\
$E_{\rm f}^{\rm SIA-octa}$ & 9.92 \cite{li2011modified} & 12.05 \cite{li2011modified} & 12.64 \cite{wang2017embedded} & 11.25 & 11.91 &  & 13.11 \cite{marinica2013interatomic}, 11.68 \cite{nguyen2006self} & 12.07 & 11.09 \\
$E_{\rm f}^{\rm vac}$ & 1.68 \cite{juslin2005analytical} & 3.52 \cite{li2011modified} & 3.49 \cite{wang2017embedded} & 3.63 \cite{mason2023empirical} & 3.30 \cite{wang2023deep} & $3.67$ \cite{rasch1980quenching} & 3.49 \cite{marinica2013interatomic}, 3.35 \cite{ma2019effect} & 3.17 & 3.36 \\ 
$E_{\rm m}^{\rm vac}$ & 1.77 \cite{li2011modified} & 1.81 \cite{li2011modified} & 1.85 \cite{chen2018new} & 1.75 \cite{mason2023empirical} & 1.95 & $1.78$ \cite{rasch1980quenching} & 1.78 \cite{nguyen2006self}, 1.73 \cite{ma2019effect} & - & 2.18 \\ 
$E_{\rm b}^{\rm divac-1NN}$ & 0.36 \cite{li2011modified} & 0.62 \cite{li2011modified} & 0.50 \cite{wang2017embedded} & 0.11 & 0.20 &  & -0.05 \cite{becquart2007ab}, 0.01 \cite{heinola2017stability} & 0.02 & 0.35 \\
$E_{\rm b}^{\rm divac-2NN}$ & 0.08 \cite{li2011modified} & 0.15 \cite{li2011modified} & 0.39 \cite{wang2017embedded} & 0.03 & -0.21 &  & -0.39 \cite{becquart2007ab}, -0.35 \cite{heinola2017stability} & -0.33 & -0.09 \\
$E_{\rm f}^{\rm surf-(100)}$ & 1.446 \cite{li2011modified} & 3.157 \cite{li2011modified} & 2.721 \cite{hao2020migration} & 3.973 \cite{mason2023empirical} & 3.900 \cite{wang2023deep} & 2.990 \cite{tyson1977surface} & 4.021 \cite{szlachta2014accuracy}, 4.149 \cite{ma2020multiscale} & 3.984 & 4.007 \\
$E_{\rm f}^{\rm surf-(110)}$ & 0.931 \cite{li2011modified} & 2.319 \cite{li2011modified} & 2.306 \cite{hao2020migration} & 3.524 \cite{mason2023empirical} & 3.302 \cite{wang2023deep} & 3.220 \cite{xu1994fourth} & 3.268 \cite{szlachta2014accuracy}, 3.396 \cite{ma2020multiscale} & 3.241 & 3.180 \\
$E_{\rm f}^{\rm surf-(111)}$ & 1.720 \cite{juslin2005analytical} & 3.222 \cite{hao2020migration} & 2.963 \cite{hao2020migration} & 4.261 \cite{mason2023empirical} & 3.563 \cite{wang2023deep} &  & 3.556 \cite{szlachta2014accuracy}, 3.829 \cite{ma2020multiscale} & 3.457 & 3.491 \\  
\hline   
Binary W-H: & & & & & & & & & \\
$E_{\rm s}^{\rm tis}$ & 1.04 \cite{wang2017embedded} & 0.86 \cite{wang2017embedded} & 1.05 \cite{wang2017embedded} & 0.80 \cite{mason2023empirical} & 0.95 \cite{wang2023deep} & $1.04$ \cite{frauenfelder1969solution} & 0.88 \cite{liu2009structure}, 0.85 \cite{heinola2010diffusion} & 0.92 & 0.92 \\
$E_{\rm s}^{\rm ois}$ & 1.40 \cite{wang2017embedded} & 1.18 \cite{wang2017embedded} & 1.40 \cite{wang2017embedded} & 1.16 \cite{mason2023empirical} & 1.35 \cite{wang2023deep} &  & 1.26 \cite{liu2009structure}, 1.33 \cite{heinola2010diffusion} & 1.30 & 1.29 \\
$E_{\rm m}^{\rm t \rightarrow t}$ & $0.34$ \cite{juslin2005analytical} & 0.23 \cite{wang2017embedded} & 0.22 \cite{wang2017embedded} & 0.22 \cite{mason2023empirical} & 0.20 \cite{wang2023deep} & 0.39 \cite{frauenfelder1969solution} & 0.20 \cite{liu2009structure}, 0.21 \cite{heinola2010diffusion} & - & 0.19 \\
$E_{\rm b}^{\rm Tplanar}$ & 0.01 \cite{wang2023deep} & 0.97 \cite{wang2023deep} & 1.04 \cite{wang2023deep} & 1.27 & 0.65 \cite{wang2023deep} & & 0.64 \cite{wang2023deep} & 0.61 & 0.56 \\
$E_{\rm b}^{\rm Oplanar}$ & - \cite{wang2023deep} & 0.67 \cite{wang2023deep} & 0.65 \cite{wang2023deep} & 0.95 & 0.87 \cite{wang2023deep} & & 0.87\cite{wang2023deep} & 0.87 & 0.74 \\
$E_{\rm b}^{\rm rock-salt}$ & -0.36 \cite{wang2023deep} & 0.85 \cite{wang2023deep} & 0.36 \cite{wang2023deep} & 1.14 & 0.50 \cite{wang2023deep} & & 0.50 \cite{wang2023deep} & 0.48 & 0.52 \\
$E_{\rm b}^{\rm surf-(100)}$ & 0.39 \cite{wang2023deep} & -2.29 \cite{wang2023deep} & -0.18 \cite{wang2023deep} & -0.31 \cite{mason2023empirical} & -0.90 \cite{wang2023deep} & & -0.87 \cite{piazza2018saturation}, -0.91 \cite{bergstrom2019hydrogen} & - & -0.80 \\
$E_{\rm b}^{\rm surf-(110)}$ & 0.09 \cite{wang2023deep} & -1.08 \cite{wang2023deep} & 0.28 \cite{wang2023deep} & -0.49 \cite{mason2023empirical} & -0.76 \cite{wang2023deep} & & -0.76 \cite{piazza2018saturation} -0.75 \cite{bergstrom2019hydrogen} & - & -0.75 \\
$E_{\rm b}^{\rm surf-(111)}$ & 0.51 \cite{wang2023deep} & -0.47 \cite{wang2023deep} & -0.01 \cite{wang2023deep} & -0.39 \cite{mason2023empirical} & -0.81 \cite{wang2023deep} & & -0.62 \cite{bergstrom2019hydrogen} & - & -0.59 \\ 
\hline   
Element H: & & & & & & & & & \\
$E_{\rm f}^{\rm H_2}$ & -4.75 \cite{wang2017embedded} & -4.75 \cite{wang2017embedded} & -4.72 \cite{wang2017embedded} & -4.24 \cite{mason2023empirical} & -4.53 \cite{wang2023deep} & -4.52 \cite{mason2023empirical} & -4.53 \cite{wang2023deep} & -4.56 & -4.59 \\
$L_{\rm b}^{\rm H_{\rm 2}}$ & 0.74 \cite{juslin2005analytical} & 0.74 \cite{li2011modified} & 0.73 \cite{wang2017embedded} & 0.74 \cite{mason2023empirical} & 0.75 \cite{wang2023deep} &  & 0.75 \cite{wang2023deep} & 0.75 & 0.74 \\
\hline\hline  % inserts single-line  
\end{tabular}
}
\label{basicproperties_results}
\end{table*}

\begin{figure*}[ht]
    \centering
    \includegraphics[width=2\columnwidth]{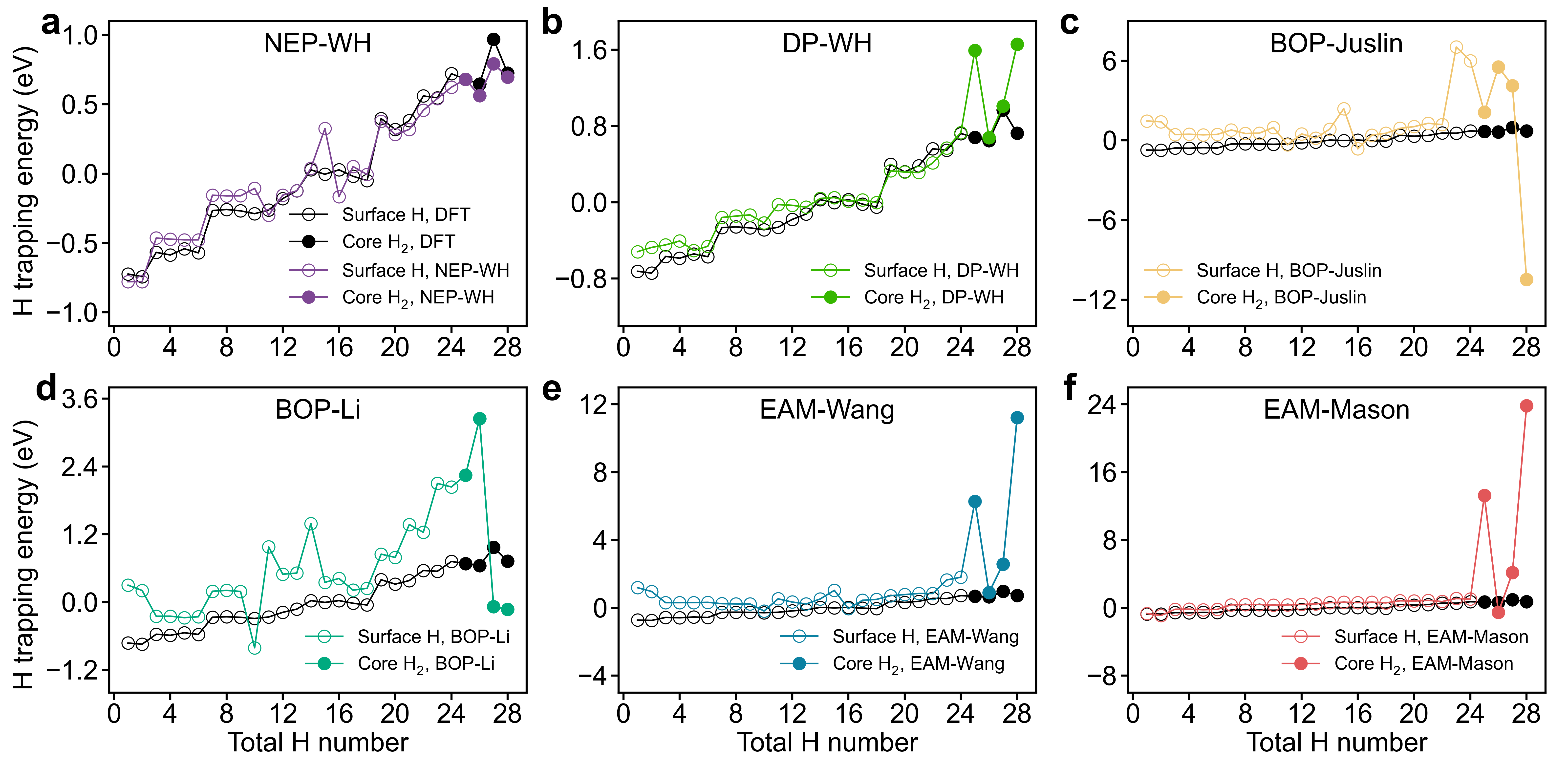}
    \caption{\textbf{Hydrogen trapping energy benchmarks.} Hydrogen trapping energy as a function of the total number of H atoms in $\rm V_{4}$ nanovoid, comparing \gls{dft} results (black circles) with predictions from (\textbf{a}) the \gls{nep}-WH model, (\textbf{b}) the \gls{dp}-WH model~\cite{wang2023deep},  (\textbf{c}) the \gls{bop}-Juslin~\cite{juslin2005analytical}, (\textbf{d}) the \gls{bop}-Li~\cite{li2011modified}, (\textbf{e}) the \gls{eam}-Wang~\cite{wang2017embedded}, and (\textbf{f}) the \gls{eam}-Mason~\cite{mason2023empirical}. Open symbols represent H adatoms on the nanovoid surface, and filled symbols indicate the formation of $\rm H_{2}$ molecules within nanovoid. The stable $\rm V_{4}H_{n}$ structures are extracted from Ref.~\cite{hou2019predictive}.}
    \label{fig:VacmHn_results}
\end{figure*}

\vspace{0.5cm}
\noindent\textbf{Evaluation of the basic properties of W-H systems}

\noindent To assess the reliability of the \gls{nep}-WH model beyond \gls{rmse} metrics, we compare its predictions for a broad range of static properties of W-H systems with those from other models, including \gls{bop} models~\cite{juslin2005analytical, li2011modified}, \gls{eam} models~\cite{wang2017embedded, mason2023empirical}, and the \gls{dp}-WH model~\cite{wang2023deep}, as well as with previous experimental and \gls{dft} results, and additional \gls{dft} calculations conducted in this work (\autoref{basicproperties_results}). The evaluated properties span elemental W and H, as well as binary W-H system. We performed static calculation for mainly using the \gls{nep}-WH model, and also other models when the relevant data are missing from the literature. The definitions and calculation details of all benchmark properties are provided in Supplementary Note S4.

For elemental W, all models reasonably predict lattice parameters and elastic constants. The formation energy of interstitial dumbbells ($E_{\rm f}^{\rm SIA}$) is correctly ordered only by the \gls{bop}-Li~\cite{li2011modified} and \gls{nep}-WH models. Monovacancy formation energies ($E_{\rm f}^{\rm vac}$) and vacancy migration barriers ($E_{\rm m}^{\rm vac}$) are well captured by all models except \gls{bop}-Juslin~\cite{juslin2005analytical}, which significantly underestimates $E_{\rm f}^{\rm vac}$. Among all models, only \gls{dp}-WH and \gls{nep}-WH correctly predict the signs of the binding energies for both first- ($E_{\rm b}^{\rm divac-1NN}$) and second-nearest neighbor divacancies ($E_{\rm b}^{\rm divac-2NN}$). For surface formation energies ($E_{\rm f}^{\rm surf}$), only \gls{dp}-WH and \gls{nep}-WH reproduce the reconstruction of the (100) free surface and both the correct magnitude and relative ordering for low-index facets, while \gls{nep}-WH additionally shows strong agreement with \gls{dft} for high-index surfaces, outperforming \gls{dp}-WH (Supplementary Figure S3). Similarly, for grain boundary formation energies across various types, both \glspl{mlp} perform significantly better than empirical potentials (Supplementary Fig. S4). To further evaluate the performance of the various models in reproducing static properties of elemental W, we calculated formation energies of interstitial dislocation loops, vacancy clusters, and the \gls{gsf} energies along the $\left \langle 111 \right \rangle$ direction of the (1 $\overline{1}$ 2) plane and (1 $\overline{1}$ 0) plane (Supplementary Figure S5). Both \gls{nep}-WH and \gls{dp}-WH exhibit good accuracy, while the BOP and EAM models are less accurate.

In the binary W-H system, all models reasonably reproduce H solution energies at both \glspl{tis} ($E_{\rm s}^{\rm tis}$) and \glspl{ois} ($E_{\rm s}^{\rm ois}$) in \gls{bcc} W. They also correctly identify the minimum-energy diffusion path as hops between adjacent \glspl{tis}, with predicted migration barriers ($E_{\rm m}^{t \rightarrow t}$) in good agreement with \gls{dft} and experimental results. Furthermore, the solution energy of hydrogen at a \gls{tis} in hydrostatically strained \gls{bcc} W lattice, and the migration barrier along the minimum energy path as a function of hydrostatical lattice strain (Supplementary Figure S5 f-g) calculated by both \gls{nep}-WH and \gls{dp}-WH exhibit good accuracy. 
Recent \gls{dft} studies reveal that hydrogen prefers energetically to form planar clustered structures, where a planar hydrogen cluster comprising \gls{tis} H or \gls{ois} H aligns along a (001) plane, rather than existing as individual interstitials. Additionally, the rock-salt mono-hydride structure, with hydrogen occupying all [001] type \gls{ois}, is more energetically stable than both \gls{tis} and \gls{ois} planar self-clusters~\cite{hou2018hydrogen, hou2020hydrogen}. Both the \gls{dp}-WH and \gls{nep}-WH models successfully reproduce the \gls{dft}-predicted energetic trend: 
$E_{\rm b}^{\rm rock\text{-}salt} <  E_{\rm b}^{\rm Tplanar} < E_{\rm b}^{\rm Oplanar} < E_{\rm s}^{\rm tis} < E_{\rm s}^{\rm ois}$, while all empirical potential fail to capture this order (Supplementary Fig. S6). Similarly, for the binding energy of a hydrogen atom at the most stable adsorption site on low-index surfaces ($E_{\rm b}^{\rm surf}$), \gls{mlp} predictions closely match \gls{dft} reference values, while all empirical potentials exhibit either qualitatively incorrect trends or large quantitative deviations. Finally, for the H\textsubscript{2} molecule, all models well accurately reproduce the formation energy $E_{\rm f}^{\rm H_2}$ and bond length $L_{\rm b}^{\rm H_2}$ of the H\textsubscript{2} molecule in vacuum. 

Across the benchmarks above, both the \gls{nep}-WH and \gls{dp}-WH models clearly outperform all considered empirical potentials in capturing W properties, and W-H interactions in vacancy-free environments, demonstrating strong consistency with \gls{dft} reference values. However, this consistency breaks down in the presence of hydrogen-vacancy complexes. The \gls{dp}-WH model has been shown to inaccurately predict the maximum number of H atoms that can be accommodated in monovacancies due to an insufficient diverse training set~\cite{wang2023deep}, raising concerns on its reliability for simulating hydrogen bubble nucleation and growth~\cite{liu2009vacancy, hou2019predictive}. To further assess these limitations, we extended our benchmarks to evaluate hydrogen trapping energetics in larger vacancy clusters ($\rm V_{4}H_{n}$, $n > 0$) at finite temperatures. The $\rm V_{4}H_{n}$ structures were obtained by sequentially introducing hydrogen atoms at \SI{600}{\kelvin}, following the protocol described in Ref.~\cite{hou2019predictive} (see Supplementary Note S3 for details). While \gls{nep}-WH continues to deliver accurate predictions, \gls{dp}-WH fails to reproduce correct hydrogen trapping energies when H\textsubscript{2} molecules form in V\textsubscript{4} clusters (\autoref{fig:VacmHn_results}). Other empirical potentials exhibit even larger deviations at all trapping stages. Currently, \gls{nep}-WH is the only W-H potential capable of reliably describing both hydrogen trapping and $\rm H_{2}$ nucleation in nanovoids.

\begin{figure}[h]
    \centering
    \includegraphics[width=1\columnwidth]{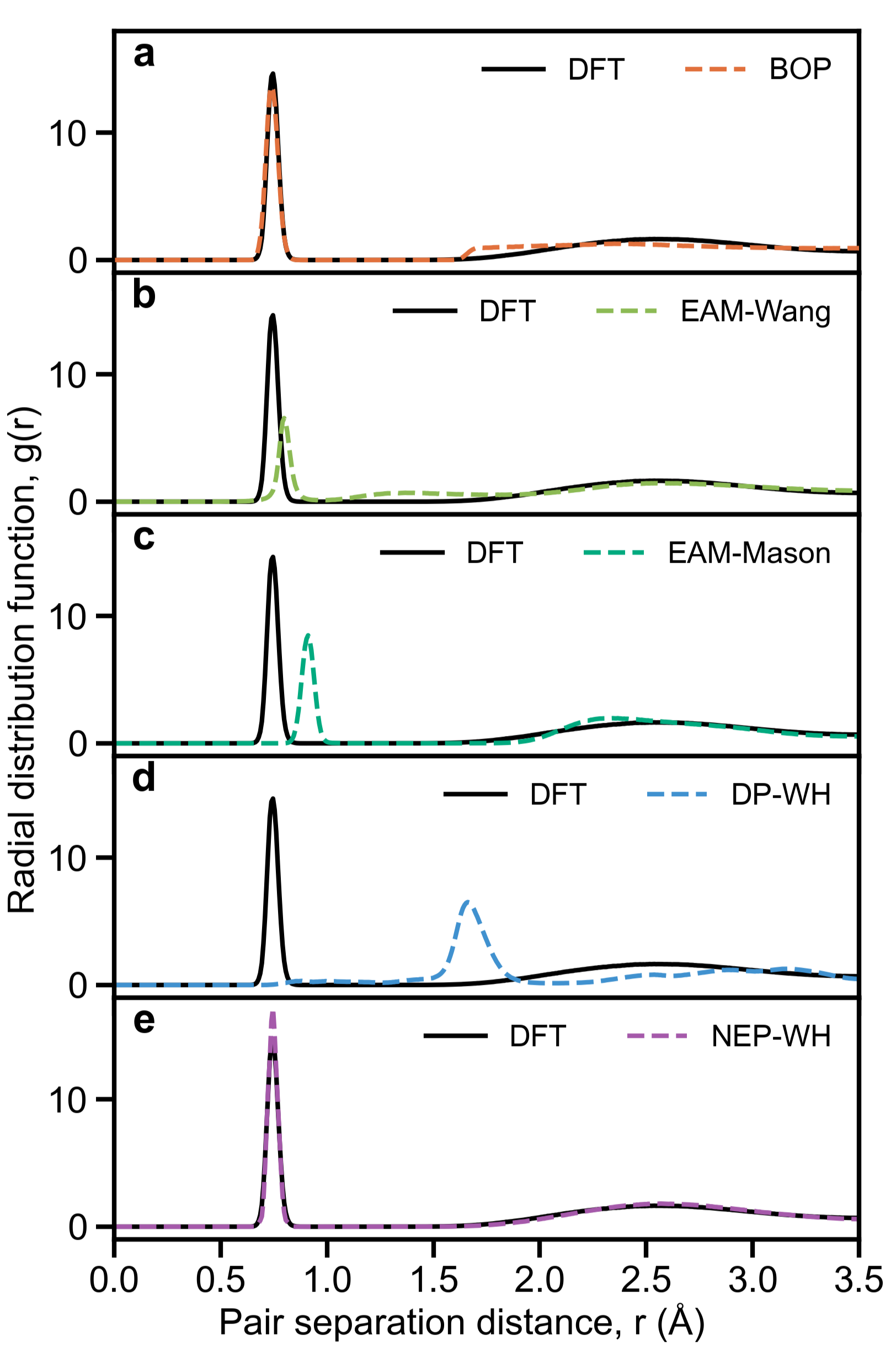}
    \caption{\textbf{Hydrogen molecule \glspl{rdf} benchmarks.} H-H \glspl{rdf} of 60 $\rm H_{2}$ systems at \SI{300}{\kelvin} and \SI{5}{\giga\pascal}, calculated using (\textbf{a}) \glspl{bop}~\cite{juslin2005analytical, li2011modified}, (\textbf{b}) \gls{eam}-Wang~\cite{wang2017embedded}, (\textbf{c}) \gls{eam}-Mason~\cite{mason2023empirical}, (\textbf{d}) \gls{dp}-WH model \cite{wang2023deep} and (\textbf{e}) \gls{nep}-WH model, benchmarked against \gls{dft}-driven \gls{md} results. All results are derived from the average of a 10 ps simulation using the NVT ensemble. Note that the two \glspl{bop} (\gls{bop}-Li~\cite{li2011modified} and \gls{bop}-Juslin~\cite{juslin2005analytical}) share identical H-H interactions terms.}
    \label{fig:rdf_results}
\end{figure}

All properties discussed above were evaluated as static, zero-pressure configurations to establish the fundamental material characteristics. However, experimental observations have estimated hydrogen bubble pressures to reach approximately \SI{10}{\giga\pascal} at \SI{300}{\kelvin} in H$_\textrm{2}$ bubbles~\cite{van1988hydrogen, shu2007blister}, underscoring the importance of accounting for pressure effects at room temperature. \autoref{fig:rdf_results} presents the H-H \glspl{rdf} at \SI{300}{\kelvin} and \SI{5}{\giga\pascal}, computed from \gls{md} simulations using various potential models and compared against \gls{dft}-driven \gls{md} results. The \gls{eam} model and \gls{dp}-WH fail to reproduce the average intramolecular H-H distance associated with the first \gls{rdf} peak, while the \gls{bop} model fails to capture the average intermolecular H-H distance reflected in the second peak. In contrast, \gls{nep}-WH model accurately captures both features, making it the only potential capable of reliably simulating bubble behaviour under intrinsic H\textsubscript{2} molecule gas pressure. Its predictive accuracy is further validated at higher pressures of \SI{10}{\giga\pascal} and \SI{20}{\giga\pascal}, as shown in Supplementary Figure S7.

\vspace{0.5cm}
\noindent\textbf{Hydrogen bubble formation and evolution}

\noindent After validating the reliability of the developed \gls{nep}-WH in capturing the lattice and defect properties of both W and W-H systems, and intrinsic H$_\textrm{2}$ bubble behavior, we investigated the behavior of hydrogen dissolution in W nanovoids of varying sizes and the size dependence of the resulting hydrogen-bubble pressure. The dissolution process was modeled via an iterative procedure, where H$_\textrm{2}$ molecules were sequentially inserted into the void core followed by structural relaxation at finite temperature (see Methods for details), effectively capturing the dynamic evolution. The bubble pressure was calculated as the hydrostatic pressure in the core sphere with a radius of \SI{2}{\angstrom} smaller than the nanovoid radius.  

\autoref{fig:bubble}a shows the evolution of hydrogen bubble pressure in nanovoids of varying radii over simulation time. For all nanovoids, H$_\textrm{2}$ molecules initially fill the voids rapidly, resulting in a swift increase in pressure. During this process, some H$_\textrm{2}$ molecules dissociate into H atoms that adsorb onto the nanovoid surface, while others remain trapped inside (see Supplementary Movie 1). A representative atomic snapshot of the void region at \SI{0.109}{\nano\second} for a \SI{1}{\nano\meter} nanovoid is shown in \autoref{fig:bubble}b, consistent with the stable dissolution configuration previously predicted by \gls{dft} calculations \cite{hou2019predictive}. As additional H$_\textrm{2}$ molecules are introduced, the system reaches equilibrium, maintaining a steady population of H$_\textrm{2}$ molecules within the nanovoid and consequently stabilizing the hydrogen bubble pressure. Due to the higher solubility of hydrogen atoms in the tungsten matrix compared to helium \cite{jiang2010strong}, and the limited void sizes considered \cite{hou2019predictive}, we did not observe the continuous pressure buildup reported for helium bubbles that leads to loop punching \cite{xie2017new}. Instead, excess hydrogen molecules dissociate at the void surface and diffuse into the surrounding matrix, resulting in a dynamic equilibrium of hydrogen bubble pressure within the void. 

The equilibrium hydrogen bubble pressures, averaged over the final \SI{1}{\nano\second}, are \SI{18.73}{\giga\pascal}, \SI{17.58}{\giga\pascal}, \SI{16.95}{\giga\pascal}, and \SI{16.01}{\giga\pascal} for nanovoids with radii of \SI{0.75}{\nano\meter}, \SI{1}{\nano\meter}, \SI{1.25}{\nano\meter}, and \SI{1.5}{\nano\meter}, respectively. For larger nanovoids with radii of \SI{1.75}{\nano\meter}-\SI{5}{\nano\meter}, hydrogen atoms were initially placed inside the voids and iterative procedure was applied to accelerate the simulation. The resulting hydrogen bubble pressure exhibits an inverse proportionality to the void radius, consistent with the functional form predicted by established empirical model~\cite{hou2019predictive} (see Supplementary Note S5 and Supplementary Figure S8). While quantitative differences are present, the predicted pressure range qualitatively agrees with available experimental values~\cite{van1988hydrogen, shu2007blister}. These deviations may originate from neglecting temperature and void surface energy effects. According to the Young-Laplace theorem \cite{Young1805essay}, hydrogen bubble pressure in nanovoids should be proportional to $\gamma/R$, where $\gamma$ is the surface energy and $R$ is the void radius. This relationship has long lacked direct experimental validation. Our study provides quantitative data of hydrogen pressures in nanovoids of varying sizes, demonstrating a strong linear relationship between pressure and $\gamma/R$ (see inset in \autoref{fig:bubble}a). This predictive relationship establishes a fundamental reference for future investigations of hydrogen bubble thermodynamics.

\begin{figure}
    \centering
    \includegraphics[width=\columnwidth]{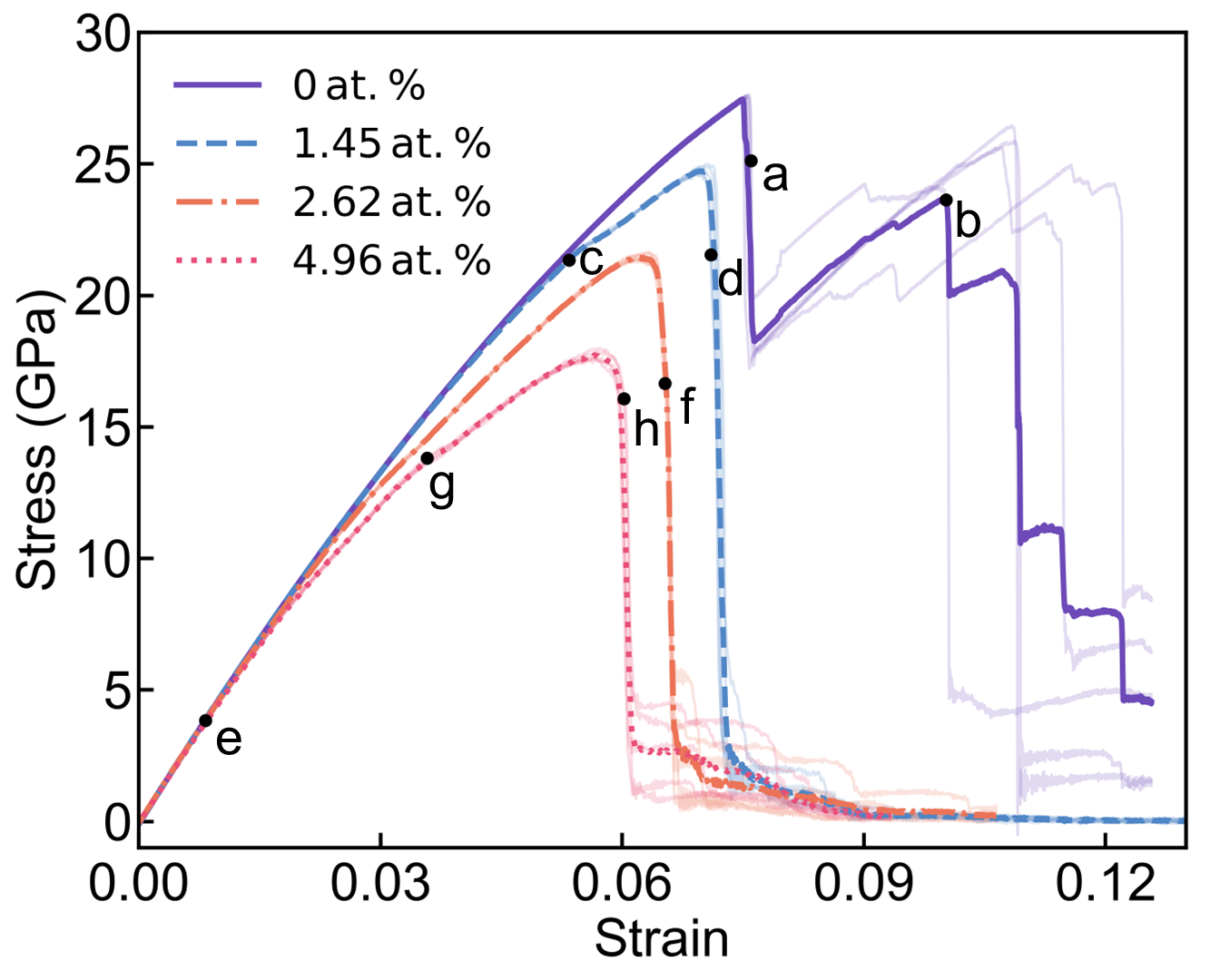}
    \caption{\textbf{Stress-strain curves for nanovoid under uniaxial tension along the [100] direction at varying hydrogen concentrations.} A \SI{1}{\nano\meter} void is examined at 0 at.\% (purple solid lines), 1.45 at.\% (blue dash lines), 2.62 at.\% (orange dash-dotted lines), and 4.96 at.\% (pink dotted lines) hydrogen concentrations, corresponding to the initial void and the void after 474, 855, and 1619 H$_\textrm{2}$ molecule insertion during hydrogen bubble simulations (\autoref{fig:bubble}), respectively. Lighter lines represents individual results from five independent simulations for each configuration.}
    \label{fig:tension}
\end{figure}

\begin{figure*}
    \centering
    \includegraphics[width=2\columnwidth]{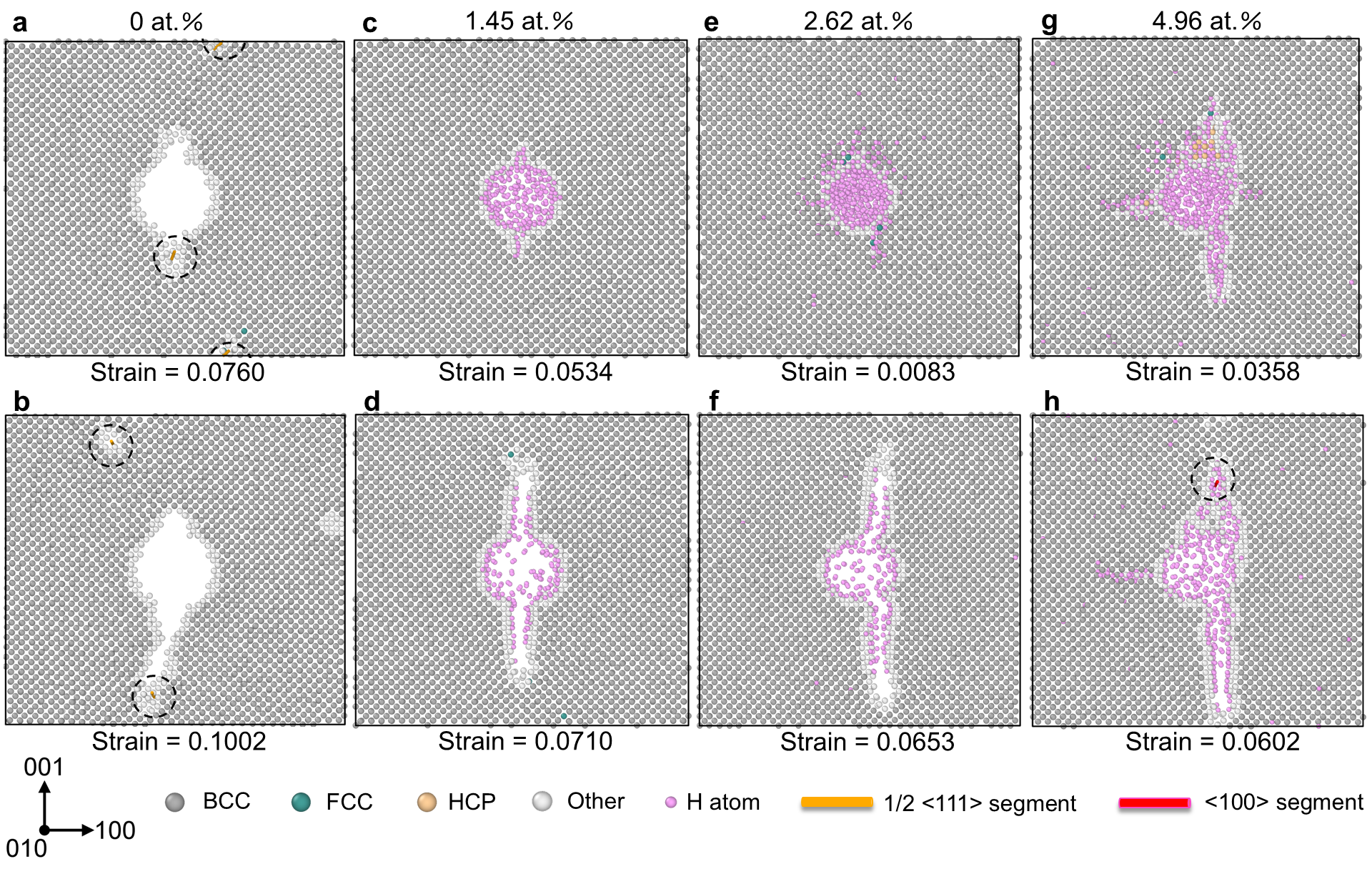}
    \caption{\textbf{Atomic configurations of a void with varying H concentrations under tensile strain along the [100] orientation.} The simulation box is spanned by the [001] and [100] directions, and the view is along the [010] direction, showing \SI{5}{\angstrom}-thick slices near the void center. \textbf{a, b} show the initial nanovoid (atoms within \SIrange{1}{6}{\angstrom} from the void center) before and after crack nucleation, respectively, while \textbf{c, d}, \textbf{e, f}, and \textbf{g, h} correspond to the voids with H concentrations of 1.45 at.\%  ([010] slices spanning \SIrange{0}{5}{\angstrom}), 2.62 at.\% (\SIrange{-3}{-8}{\angstrom}), and 4.96 at.\% (\SIrange{-2}{-6}{\angstrom}), respectively. W atoms identified by the \gls{cna} method as \gls{bcc}, \gls{fcc}, \gls{hcp}, and Other are shown as large gray, green, orange, and white spheres, respectively. Pink spheres represent H atoms. Orange lines indicate 1/2 $\left \langle 111 \right \rangle$ dislocation loop segments, and red lines denote $\left \langle 100 \right \rangle$ segments.}
    \label{fig:cna}
\end{figure*}

As the hydrogen bubble approaches equilibrium, additional H$_\textrm{2}$ molecules dissociate into H atoms that diffuse into the surrounding matrix. Notably, hydrogen diffusion around all studied voids is anisotropic, with a clear preference for distribution along the \{100\} crystallographic planes. \autoref{fig:bubble}c presents an \gls{md} snapshot at \SI{1.417}{\nano\second}, illustrating the early-stage formation of platelet-like self-clusters along the \{100\} planes surrounding a \SI{1}{\nano\meter} void. Similar visualizations along the [010] and [001] directions are provided in Supplementary Figure~S9. These planar hydrogen self-clusters stochastically form and grow along the (100), (010), and (001) planes. In each orientation, the hydrogen atoms are confined to a \SI{4}{\angstrom}-thick slab spanning two adjacent tungsten atomic planes and the interstitial plane between them. During the simulations, hydrogen atoms frequently hop between these planes but rarely escape from the planar cluster. These \{100\} planar hydrogen clusters extend outward from the void surface, and the hydrogen atoms within them occupy \glspl{tis}. Their presence increases the interplanar spacing between adjacent W layers and induces \gls{bcc}-to-\gls{fcc} transitions in small, localized regions of the lattice; a minor \gls{hcp} phase also emerges at the intersections of the (100) and (001) planar hydrogen clusters. 

With ongoing hydrogen diffusion, the planar clusters along \{100\} planes expand outward from the void surface. Although atomic positions of hydrogen fluctuate slightly, the cluster thickness and orientation remain unchanged. This expansion enhances the extent of the \gls{fcc} phase transformation. However, these \gls{fcc} domains are dynamically unstable, repeatedly forming and annihilating during the simulation. This instability likely arises from the severe lattice distortions required for the \gls{bcc}–\gls{fcc} transformation in W, which leads to the absence of the rock-salt structure in our \gls{md} simulations and is in agreement with prior results~\cite{wang2023deep}. At the same time, the intersections of planar hydrogen clusters along the \{100\} planes promote the formation of \gls{hcp} phases that act as new hydrogen traps (see \autoref{fig:bubble}d). Two wedge-shaped hydrogen-rich \gls{hcp} regions gradually develop near the void, reaching local hydrogen concentrations of 1.3-1.5 H atoms per W atom. Both the planar clusters and the hydrogen-rich \gls{hcp} regions continue to grow until the end of the simulation (\autoref{fig:bubble}(d)). 

Hydrogen self-clustering has previously been reported under conditions of high hydrogen concentration or in the presence of local stress fields \cite{smirnov2018stress,hou2020hydrogen, wang2023deep}. Hydrogen-rich \gls{hcp} regions have also been observed near cracks under strained Fe \cite{song2013atomic}. Here, we directly observe pronounced hydrogen aggregation along \{100\} planes and hydrogen-rich \gls{hcp} regions near bubble surfaces. We attribute this behavior to the persistent internal H$_\textrm{2}$ bubble pressure and the H dissociation around bubble surface. This directional aggregation may significantly influence the mechanical response of W, underpinning the experimentally observed intragranular blisters that preferentially develop along the \{100\} planes \cite{guo2018edge,chen2020growth,chen2020nucleation}. In the following section, we investigate the mechanical response of these hydrogen-enriched configurations.

\vspace{0.5cm}
\noindent\textbf{Uniaxial tension}

\noindent To assess the impact of hydrogen bubble evolution on the mechanical properties of tungsten, we analyzed the uniaxial tensile response of representive configurations sampled from the hydrogen bubble simulation. Specifically, we examined four cases: the initial void (hydrogen concentration $C_{\rm H}$ = 0 at.\%) and hydrogen bubbles with $C_{\rm H}$ = 1.45 at.\% (\autoref{fig:bubble}b), $C_{\rm H}$ = 2.62 at.\% featuring \{100\} planar clusters (\autoref{fig:bubble}c), and $C_{\rm H}$ = 4.96 at.\% containing both planar clusters and hydrogen-rich \gls{hcp} regions (\autoref{fig:bubble}d). Uniaxial tensile simulations were conducted along the [100] orientation at \SI{600}{\kelvin} with a strain rate of \SI{1e8}{\per\second}. As shown in \autoref{fig:tension}, the pristine void without hydrogen exhibits the highest tensile strength of \SI{27.46 \pm 0.01}{\giga\pascal}. Introducing hydrogen significantly reduces the strength, decreasing to \SI{24.73 \pm 0.07}{\giga\pascal} at $C_{\rm H}$ = 1.45 at.\%, \SI{21.44 \pm 0.04}{\giga\pascal} at $C_{\rm H}$ = 2.62 at.\%, and \SI{17.73 \pm 0.06}{\giga\pascal} at $C_{\rm H}$ = 4.96 at.\%. Meanwhile, the fracture behavior transitions from ductile to brittle, with the ultimate strain dropping from $\sim$0.12 in the hydrogen-free void to $\sim$0.06 at the highest hydrogen concentration. These results highlight the pronounced embrittlement of tungsten induced by hydrogen bubbles.

Pristine voids without hydrogen exhibit distinct dislocation emission and slip behavior under tensile strain (\autoref{fig:cna}a-b). During yielding, a pie-shaped stress concentration forms around the void center, oriented perpendicular to the tensile axis, accompanied by partial \gls{bcc} to \gls{fcc} phase transitions at its tips (Supplementary Figure S10a). Simultaneously, two 1/2 $\left \langle 111 \right \rangle$ dislocation loop fragments are emitted from the nanovoid. As the trailing ends of these dislocations exit the void, vacancy clusters form on the void surface (see \autoref{fig:cna}a and Supplementary Figure S10b). The cross-slip and recombination of dislocations initiate an additional hardening stage at a strain range of $\sim$0.0760 - $\sim$0.1002 (\autoref{fig:tension}) and leave behind some vacancies along the slip path (Supplementary Figure S10b-c). Continued dislocation emission leads to coalescence of these vacancy clusters, culminating in fracture (\autoref{fig:cna}b). The resulting fracture surface varies slightly among independent simulations (Supplementary Figure S11b-f), shaped by dislocation slip and vacancy distribution.

The presence of hydrogen significantly suppresses dislocation emission and promotes premature fracture. Under uniaxial tension, hydrogen bubbles undergo pronounced brittle cleavage (\autoref{fig:cna}c-h). As the hydrogen concentration increases, the fracture crack evolves from a symmetric (\autoref{fig:cna}d) to increasingly asymmetric (\autoref{fig:cna}f,h) about the vertical axis of nanovoid, which is perpendicular to the tensile direction. This hydrogen concentration-dependent crack morphology is closely linked to the formation and evolution of planar hydrogen clusters during tensile loading.

In the nanovoid containing 1.45 at.\% hydrogen, no planar hydrogen clusters are present initially. However, under uniaxial tension, hydrogen atoms migrate toward pie-shaped stress concentration region and progressively assemble into planar clusters along the (100) crystallographic planes, as shown at a representative strain of 0.0534 prior to fracture (\autoref{fig:cna}c and Supplementary Figure S10e). As strain increases, the clusters expand and further separate adjacent tungsten layers. As a result, cracks nucleate at the nanovoid surface and propagate outward. Hydrogen atoms preferentially segregate to the crack surfaces, and some hydrogen molecules diffuse into the crack, collectively promoting brittle cleavage. The resulting fracture surface aligns along the (100) plane, exhibiting a symmetric fracture crack about the vertical axis of the nanovoid center (\autoref{fig:cna}d and Supplementary Figure S11h-i). 

\begin{figure}
    \centering
    \includegraphics[width=\columnwidth]{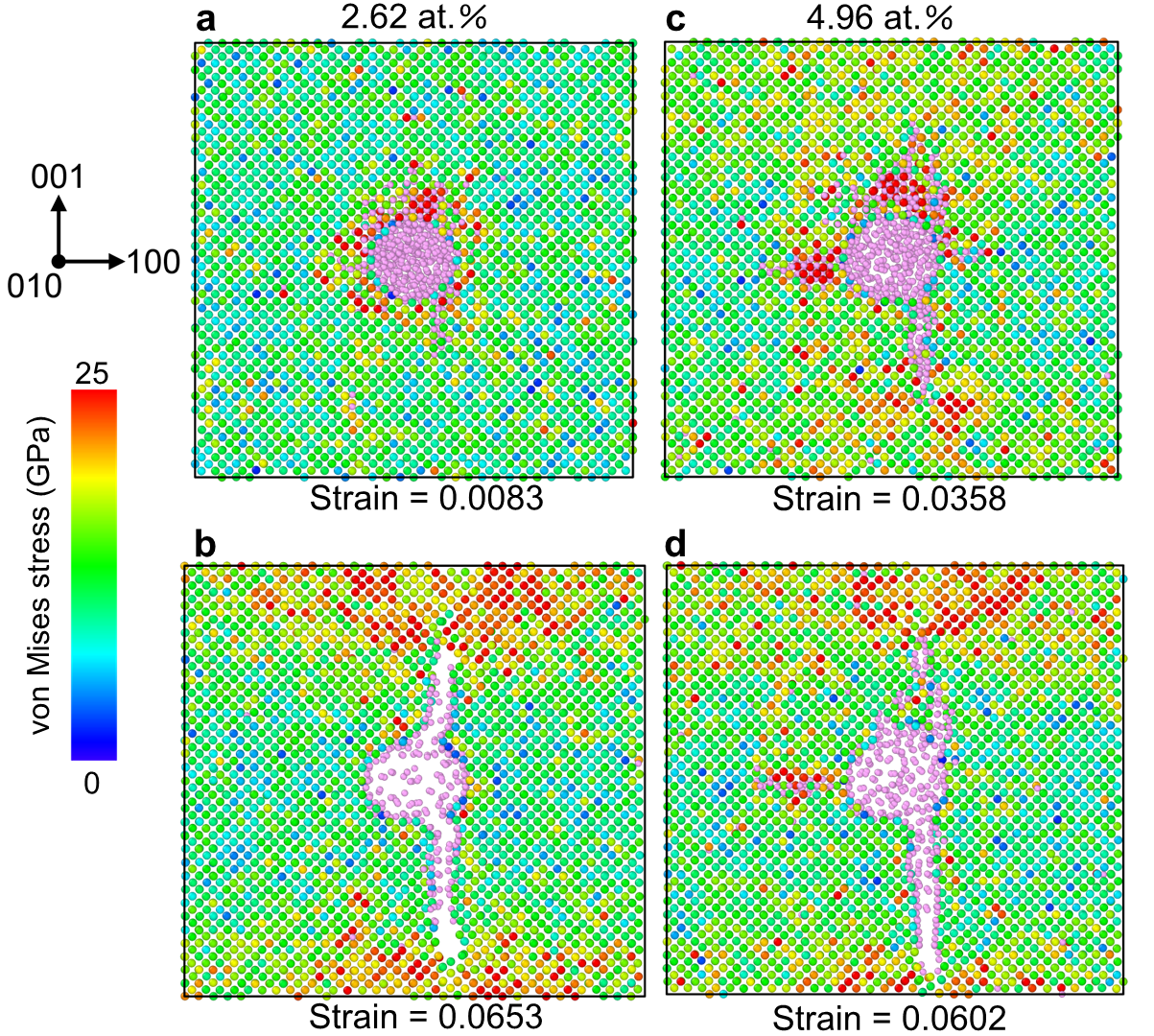}
    \caption{\textbf{Profiles of von Mises stress in the tungsten \SI{1}{\nano\meter} voids with varying H concentration under uniaxial tension along the [100] direction.} \textbf{a, b} show configurations before and after crack nucleation for a void with $C_{\rm H}$ = 2.62 at.\%, featuring \{100\} planar clusters, respectively. \textbf{c, d} show corresponding configurations for $C_{\rm H}$ = 4.96 at.\%, containing both planar clusters and hydrogen-rich \gls{hcp} regions. The view is taken along the [010] direction, showing \SI{5}{\angstrom}-thick slices near the (010) planer H cluster region. W atoms are colored by stress, and H atoms are shown in pink.}
    \label{fig:misesstress}
\end{figure}

A more pronounced hydrogen-induced ductile-to-brittle transition is observed in nanovoids with higher hydrogen concentrations (\autoref{fig:cna}e-h). In both the 2.62 at.\% and 4.96 at.\% cases, hydrogen platelets are initially distributed along all three principal crystallographic directions (see \autoref{fig:bubble}c-d and Supplementary Figure S9). Under uniaxial tension along the [100] direction, cracks preferentially propagate along the (100) hydrogen cluster planes rather than through the void center. The formation of hydrogen platelets generates localized stress concentrations (\autoref{fig:misesstress}a,c), which further accelerate brittle cleavage and reduce the ultimate tensile strength of the material (\autoref{fig:misesstress}b,d). During this process, hydrogen atoms dissolved from other hydrogen-rich regions—such as (001) and (010) planar clusters—migrate toward and accumulate along the newly formed crack surfaces (see Supplementary Movie 2 and Movie 3), further promoting fracture. Specifically, in the 2.62 at.\% case, this leads to an asymmetric fracture crack about the vertical axis of the nanovoid center (\autoref{fig:cna}f and Supplementary Figure S11n-r).

Compared to the 2.62 at.\% case, where only planar hydrogen clusters are present, the 4.96 at.\% case exhibits brittle cleavage in voids where both hydrogen-rich \gls{hcp} region and planar hydrogen clusters coexist (\autoref{fig:cna}g). In this case, the initial (100) hydrogen clusters are larger than those in the 2.62 at.\% case (see \autoref{fig:bubble}c-d), resulting in a lower tensile strength (\autoref{fig:tension}). Under uniaxial tension, the \gls{hcp} phase gradually depletes, generating vacancies and passivating the crack in that region (\autoref{fig:cna}h). During the depletion of the \gls{hcp} phase, $\left \langle 100 \right \rangle$ dislocation fragments are emitted, which absorb hydrogen atoms from the hydrogen-rich \gls{hcp} region and relieve local stress (\autoref{fig:misesstress}c,d). Notably, this dislocation emission is not associated with ductile behavior. Once new planar hydrogen clusters form within the hydrogen-rich regions, brittle cleavage is reinitiated. The interplay between dislocation emission and fracture along hydrogen platelets ultimately leads to a highly asymmetric fracture crack with respect to the vertical axis of the nanovoid center, with the crack front propagating predominantly downward, as shown in \autoref{fig:cna}(h) and Supplementary Figure S11t-x. 

\vspace{0.5cm}
\noindent{\textbf{DISCUSSION}}
\noindent In summary, we present a machine-learned \gls{nep} model for the W-H system, enabling accurate and efficient atomistic simulations of hydrogen trapping and bubble evolution in tungsten. The model reproduces \gls{dft} results across a wide spectrum of properties, including short-range interactions, lattice and defect characteristics, hydrogen energetics in both pristine and defected environments, and H$_\textrm{2}$ properties under pressure. It also captures hydrogen trapping energies in nanovoids with near-\gls{dft} fidelity, even in cases where H$_\textrm{2}$ molecules form inside, thereby enabling predictive simulations of hydrogen bubble nucleation and evolution dynamics. Such breadth of accuracy, unmatched by existing empirical or \glspl{mlp}, establishes the \gls{nep}-WH model as a significant advance. Beyond accuracy, the model delivers exceptional computational efficiency, supporting million-atom simulations on a single desktop GPU. This unique combination of fidelity and scalability offers a powerful framework for investigating hydrogen bubble dynamics with near first-principles precision.

Our hydrogen bubble simulations driven by the \gls{nep}-WH model reveal the atomic-scale mechanisms of bubble formation and evolution in nanovoids. H$_\textrm{2}$ molecules preferentially occupy the void core, while $\rm H$ accumulates on the void surface. H$_\textrm{2}$ molecules aggregation induces a rapid pressure increase that equilibrates under continuous hydrogen injection. Excess hydrogen diffuses to form planar clusters at \glspl{tis} along the \{100\} planes, and their intersections yield hydrogen-rich \gls{hcp} regions at high concentrations. 

Uniaxial tensile simulations reveal a ductile-to-brittle transition, driven by planar hydrogen clusters that suppress dislocation emission from void surface and promote brittle cleavage. The fracture response depends strongly on hydrogen concentration: (i) at low concentrations, symmetric fracture surfaces pass through the nanovoid center, as planar hydrogen clusters form dynamically at the crack front; (ii) at medium concentrations, pre-existing planar hydrogen clusters guide fracture along their planes, producing crack surfaces aligned with the cluster orientation; (iii) at high concentrations, the interplay between dislocation emission from H-rich \gls{hcp} regions and fracture of planar hydrogen clusters generates a highly asymmetric fracture morphology, with the crack front propagating unidirectionally.

Recent experiments on tungsten exposed to low-energy hydrogen plasma have reported intragranular blisters preferentially developing on {100} planes \cite{guo2018edge,chen2020growth,chen2020nucleation}, consistent with our \gls{md} observations. This correlation suggests that the experimentally observed crack-shaped blisters may originate from {100} planar hydrogen clusters surrounding voids. We propose that once the internal pressure of hydrogen bubbles exceeds a critical threshold, the combined effect of bubble pressure and planar cluster-induced stress concentration initiates cracks. These cracks subsequently propagate by incorporating nearby H, ultimately forming cavities observed experimentally. Together, these mechanistic insights establish a direct connection between atomistic bubble evolution and macroscopic blistering, providing guidance for the design of plasma-facing tungsten components in fusion energy applications. 

\vspace{0.5cm}

\noindent{\textbf{METHODS}}

\noindent{\textbf{Density functional theory calculations}}

\noindent All \gls{dft} calculations were performed using \textsc{vasp} code~\cite{kresse1996efficiency,kresse1996efficient} with exchange-correlation described by the generalized gradient approximation proposed by Perdew-Wang (PW91)~\cite{perdew1992atoms}. The Brillouin zone is sampled by the Monkhorst-Pack scheme with a grid spacing of \SI{0.2}{\per\angstrom}. The projector-augmented-wave method was applied, and a plane-wave cutoff energy of \SI{500}{\electronvolt} was used. The 5$d$ and 6$s$ electrons of W and the 1$s$ electron of H were treated as valence electrons. Convergence thresholds were set to \SI{1e-5}{\electronvolt} and \SI{1e-2}{\electronvolt\per\angstrom} respectively. The \texttt{INCAR} input file for \textsc{vasp} package is provided in Supplementary Note~S2.

\vspace{0.5cm}

\noindent{\textbf{Machine learning potentials}}

\noindent We employ the \gls{nep} approach~\cite{fan2021neuroevolution, song2024general} to construct the \gls{nep}-WH model. The \gls{nep} approach provides a framework for generating highly efficient \glspl{mlp} and has been successfully applied to model structural, chemical, thermal, mechanical, and transport properties of complex materials~\cite{ying2025advances}. Specifically, we adopt the latest NEP4 implementation, which has been applied to develop a general-purpose potential for many metals and their alloys~\cite{song2024general}.

The name \gls{nep} originates from its two key components. On the one hand, it is a \gls{nn} potential model~\cite{behler2007generalized}; on the other hand, the free parameters in the model are trained using an evolutionary algorithm~\cite{fan2021neuroevolution}. The input layer of the \gls{nn} consists of a number of descriptors $\mathbf{q}_i$, which represent the local chemical environments of a central atom $i$. There are two types of descriptors used in NEP4: radial descriptors $q_n^i$ that depend only on interatomic distances, and angular descriptors $q_{nl}^i$ that depend on both atom distances and bond angles. The index $n$ in $q_n^i$ ranges from 0 to $n_{\rm max}^{\rm R}$ and the index $n$ in $q_{nl}^i$ ranges from 0 to $n_{\rm max}^{\rm A}$. 
The index $l$ ranges from 0 to $l_{\rm max}$. To balance accuracy with efficiency, we choose $n_{\rm max}^{\rm R}=9$, $n_{\rm max}^{\rm A}=8$, and $l_{\rm max}=4$.
The radial and angular descriptors are smoothly cut at \SI{6}{\angstrom} and \SI{5}{\angstrom}, respectively. At short distances, the \gls{nep} model is smoothly connected to  the \gls{zbl} potential~\cite{ziegler1985stopping} to account for the strong repulsive forces encountered under extreme conditions~\cite{liu2023large}.

The \gls{nn} model takes the descriptors as the input layer neurons and output the site energy $U_i$ of the atom $i$. With a hidden layer with $N_{\rm neu}$ neurons, the site energy can be explicitly written as 
\begin{equation}
\label{equation:Ui}
U_i = \sum_{\mu=1} ^{N_\mathrm{neu}} w ^{(1)} _{\mu} \tanh\left(\sum_{\nu=1} ^{N_\mathrm{des}} w ^{(0)}_{\mu\nu} q^i_{\nu} - b^{(0)}_{\mu}\right) - b^{(1)}.
\end{equation}
Here, $\tanh(x)$ is the activation function, $w^{(0)}$ are the weight parameters connecting the input layer (with dimension $N_{\rm des}$) and the hidden layer, $w^{(1)}$ represents the weight parameters connecting the hidden layer and the output layer (the site energy), $b^{(0)}$ represent the bias parameters in the hidden layer, and $b^{(1)}$ is the bias parameter in the output layer. We use a relatively large value of $N_{\rm neu}=120$ in our model. All these parameters are optimized by minimizing a loss function that is a weighted sum of the \glspl{rmse} of energy, force, and virial. Details of the \texttt{nep.in} and \texttt{zbl.in} input files are provided in Supplementary Note~S3.

\vspace{0.5cm}
\noindent{\textbf{Molecular dynamics}} 

\noindent The \gls{nep}-WH model was used to conduct large-scale \gls{md} simulations to investigate hydrogen solubility in nanovoid and the influence of H on the mechanical properties of W. All simulations were performed using \textsc{gpumd} package~\cite{xu2025gpumd}, and visualizations were carried out with the \textsc{ovito} software~\cite{stukowski2009visualization}.

We first constructed a cubic simulation cell with dimensions of $10 \times 10 \times 10$ nm$^3$ along the [100], [010], and [001] directions. Periodic boundary conditions were applied in all three directions. Nanovoids with radii of \SI{0.75}{\nano\meter}, \SI{1}{\nano\meter}, \SI{1.25}{\nano\meter}, \SI{1.5}{\nano\meter}, \SI{1.75}{\nano\meter} were created at cell center. For larger void radii of \SI{2}{\nano\meter}, \SI{3}{\nano\meter}, \SI{4}{\nano\meter}, and \SI{5}{\nano\meter}, the domain size was increased while maintaining a constant void volume fraction of 0.014, in order to minimize periodic boundary effects.  An initial energy minimization was performed to relax the atomic configurations and eliminate any residual forces. The W system was then equilibrated for \SI{20}{\pico\second} in the isothermal–isobaric (NPT) ensemble at \SI{600}{\kelvin}, with a time step of \SI{2.0}{\femto\second}. H$_\textrm{2}$ molecules were randomly introduced into the void cores, with each pair of H atoms seperated by more than \SI{1.3}{\angstrom}. The void core was defined as a sphere with a radius of \SI{2}{\angstrom} smaller than the actual void radius, to avoid interaction between H and the W surface. After every 10 H$_\textrm{2}$ molecules were inserted, the system was relaxed for \SI{1}{\pico\second} in the canonical (NVT) ensemble, with a time step of \SI{0.4}{\femto\second}. If an insertion attempt failed, it was skipped. Additionally, a \SI{10}{\pico\second} relaxation was performed after every 100 insertions to ensure full structural equilibration.

To study the effect of H on mechanical behavior of W, uniaxial tension was applied along the [100] direction for systems containing a \SI{1}{\nano\second} void with varying H concentrations, with a strain rate of \SI{1e-8}{\per\second}. Simulations were carried out in the canonical (NVT) ensemble with a time steps of \SI{0.4}{\femto\second} W-H systems and \SI{2}{\femto\second} for pure W systems.

\vspace{0.5cm}

\noindent{\textbf{DATA AVAILABILITY}}

\noindent The \gls{nep}-WH potential files are freely available at the Zenodo repository \url{https://doi.org/10.5281/zenodo.16973981}. The corresponding reference datasets will be made available upon acceptance of this work. Additional data that support the findings of this study are available from the corresponding author upon reasonable request.

\vspace{0.5cm}

\noindent{\textbf{\large Author contributions}}
\noindent P.Q. initiated and conceived the study. Y.B. and K.S. prepared the training and test structures. Y.B. performed the DFT calculations. Y.B. and J.L. tested the various hyperparameters and trained the NEP model. P.Y. analyzed the descriptor space. Y.B. and Y.N. evaluated the NEP model's performance on basic properties. Y.B., K.S., and Y.W. performed MD simulations. P.Q. and P.Y. supervised the project. All authors discussed the results and contributed to the writing of the paper.

\vspace{0.5cm}

\noindent{\textbf{\large Declaration of competing interest}}
\noindent The authors declare that they have no competing interests.

\vspace{0.5cm}

\noindent{\textbf{\large Acknowledgement}}
\noindent This research was funded by the Advanced Materials-National Science and Technology Major Project (No.2024ZD0607800), the National Key Research and Development Program of China (2023YFB3506704), the National Natural Science Foundation of China (52371003), and the University of Science and Technology Beijing High Performance Computing university-level public platform. P.Y. is supported by the Israel Academy of Sciences and Humanities \& Council for Higher Education Excellence Fellowship Program for International Postdoctoral Researchers. Y.W. is supported by the Academy of Finland's Grant No. 353298 under the European Union – NextGenerationEU instrument.

\vspace{0.5cm}

\noindent{\textbf{\large References}} 
\bibliography{ref.bib}

\end{document}